\documentstyle[12pt,epsf]{article}
\newcommand{\newsection}{    % Numeration of eqs. is automatic
\setcounter{equation}{0}
\section}
\def\appendix#1{
\addtocounter{section}{1}
\setcounter{equation}{0}
\renewcommand{\thesection}{\Alph{section}}
\section*{Appendix \thesection\protect\indent #1}
\addcontentsline{toc}{section}{Appendix \thesection\ \ \ #1}
}
\newcommand{\rf}[1]{(\ref{#1})}

\def\be{\begin{equation}}
\def\ee{\end{equation}}
\newcommand{\beq}{\begin{equation}}
\newcommand{\eeq}{\end{equation}}
\newcommand{\bea}{\begin{eqnarray}}
\newcommand{\eea}{\end{eqnarray}}

\newcommand{\om}{\omega}

\newcommand{\tr}{{\,\rm tr}\:}

\begin{document}
\topmargin 0pt
\oddsidemargin 5mm
\headheight 0pt
\headsep 0pt
\topskip 9mm

\hfill NORDITA-96/55P 

\addtolength{\baselineskip}{0.20\baselineskip}
\begin{center}
\vspace{26pt}
{\large \bf {Phase Structure  \\ of the $O(n)$ Model on a Random Lattice 
for $n>2$}}
\newline

 \vspace{26pt}
{\sl B.\ Durhuus}\\                                           
\vspace{6pt} 
Mathematics Institute, University of Copenhagen\\ 
Universitetsparken 5, DK-2100 Copenhagen \O, Denmark\\

\vspace{18pt}
 {\sl C. Kristjansen}\\ 
 \vspace{6pt}
 NORDITA \\
  Blegdamsvej 17,
 DK-2100 Copenhagen \O, Denmark \\
 \end{center}
\vspace{20pt} 
\begin{center}
{\bf Abstract}
\end{center}
We show that coarse graining arguments invented for the analysis of multi-spin 
systems on a randomly triangulated surface apply also to the $O(n)$ model on 
a random lattice. These arguments imply that if the model has a
critical point with diverging string susceptibility, then either
$\gamma=+\frac{1}{2}$ or there exists a dual critical point with
negative string susceptibility exponent, $\tilde{\gamma}$, related to
$\gamma$ by
$\gamma=\frac{\tilde{\gamma}}{\tilde{\gamma}-1}$. Exploiting the exact
solution of the $O(n)$ model on a random lattice we show that both
situations are realized for $n>2$ and that the possible dual pairs of
string susceptibility exponents are given by
$\left(\tilde{\gamma},{\gamma}\right)=\left(-\frac{1}{m},\frac{1}{m+1}\right)$,
$m=2,3,\ldots$.
 We also show that at the critical points with positive
string susceptibility exponent
 the average number of loops on the surface diverges while the average length
of a single loop stays finite.

\noindent
\newpage

\newsection{Introduction}

Two-dimensional gravity coupled to conformal matter fields of central charge
$c\leq 1$ is a reasonably well understood subject both in the continuum
approach~\cite{KPZ88,Dav88,DK89} and in the framework of dynamical
triangulations, see~\cite{Amb94} for reviews. The case $c>1$ remains,
however, to a large extent ununderstood. In the continuum this is
reflected e.g. by the fact that critical exponents become non-real
(see, however,~\cite{Gervais}) whereas in the discrete approaches the
naive regularized models are generally not exactly solvable and most
results rely upon numerical simulations~\cite{numerical}. In~\cite{Dur94}
a kind of dual relationship between a class of discretized $c>1$
theories, or rather models with diverging susceptibility, e.g. multiple
Ising spin models, and corresponding $c<1$ models was proposed on the
basis of universality considerations. For unitary models this leads to
useful restrictions on the values of critical exponents~\cite{Dur94}.

 It is well known 
that the $O(n)$ model on a random triangulation~\cite{Kos89}
for suitable rational values of
the parameter $\nu\in ]0,1[$, related to $n$ by $n=2\cos(\nu \pi)$, 
represents (sectors of) two-dimensional gravity coupled to minimal
conformal field theories of central charge $c<1$~\cite{Kos89,DK88,KS92,EZ92}.
Recently the exact solution of the model,
valid for all values of $n$, has been 
found~\cite{EK95,EK96}.
Although it is not clear that the model for $n>2$ can reasonably be 
considered as
representing $c>1$ conformal field theories coupled to two-dimensional
gravity it is clearly of interest to investigate its properties beyond
the treshold $n=2$ where critical points with diverging susceptibility
are expected to exist. The purpose of this note is to point out that
the $O(n)$ model (at least for integer $n$) falls within the class of
models to which the arguments in~\cite{Dur94} apply and to
exploit the exact solution in~\cite{EK95,EK96} to evaluate the dual
pairs of susceptibility exponents. We find that for $n>2$
the model is either in
a branched polymer phase or that the pairs of exponents are given by
$\left(-\frac{1}{m},\frac{1}{m+1}\right)$, where $m=2,3,\ldots$.
This result in turn corroborates the universality
assumption of~\cite{Dur94} and thereby indirectly the existence of a
continuum limit at the points with diverging susceptibility. 
We show that at these points the average number of loops on the
surfaces diverges while the average length of a single loop stays finite as
suggested in an early paper by I.\ Kostov~\cite{Kosrev}.
The
investigation of the continuum limit is, however, beyond the scope of
this paper.

We start in section~\ref{presentation} by introducing the $O(n)$
model on a random lattice and present in section~\ref{coarse}
a version of the coarse graining argument of~\cite{Dur94}
valid for this model. I section~\ref{solution} we recall a few
important characteristics of the exact solution of the model and in
section~\ref{critical}, using the exact solution, we determine the
possible values of the string susceptibility exponent for
$n>2$. Section~\ref{loop} is devoted to the calculation of the average
number of loops on the surfaces and finally section~\ref{discussion}
contains some concluding remarks.

\newsection{The $O(n)$ model on a random
triangulation \label{presentation}}

\begin{figure}
\centerline{ \epsfbox{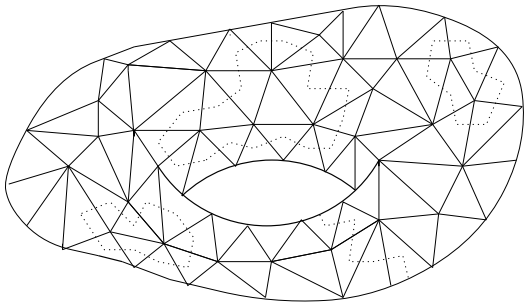}}
\caption{A surface contributing to $F_1(\underline{g})$}
\label{F0}
\end{figure}

In matrix model language the partition function of the $O(n)$ model on
a random triangulation is defined as (see~\cite{Kos89})
\beq
Z(\underline{g})
=\int_{N\times N} dM \prod_{i=1}^n dA_i\exp
\left\{-N \tr\left(\frac{1}{2}M^2+V(M)+\left(g_0M+\frac{1}{2}\right)
\sum_{i=1}^nA_i^2\right)\right\}
\label{Z}
\eeq
where $M$ and $A_1,\ldots,A_n$ are hermitian $N\times N$ matrices and
\beq
V(M)=\sum_{j=3}\frac{g_j}{j}M^j
\eeq
is a polynomial potential depending on the coupling constants
$g_3,g_4,\ldots$ and we use the collective notation $\underline{g}=
\left(g_0,g_3,g_4,\ldots\right)$. By expanding the
non-quadratic part of the exponent in powers of $g_0,g_3,g_4,\ldots$
we obtain an interpretation of $Z(\underline{g})$ as the
partition function of a gas of $n$ species of self-avoiding and
non-intersecting loops living on a random triangulation (see~\cite{Kos89}).
Specifically, after
a suitable normalization of the measures $dM$ and $dA_i$ we have
\beq
F(\underline{g})=\frac{1}{N^2}\log Z(\underline{g})=
\sum_{h=0}^{\infty} N^{-2h}F_h(\underline{g})
\eeq
where
\beq
F_h(\underline{g})=\sum_{\tau\in T_h}
\sum_{L_1,\ldots,L_n}C_{\tau}^{-1}(L_1,\ldots,L_n)
\prod_{j\geq 3} (-g_j)^{N_j(\tau)}
\left(\frac{g_0}{g_3}\right)^{|L_1|+\ldots+|L_n|}.
\label{Fh}
\eeq
Here $T_h$ denotes the set of two-dimensional closed, connected
complexes
of genus $h$ obtained by gluing together $N_j(\tau)$, $j\geq 3$, $j$-gons
along pairs of links and $L_i$, $i=1,\ldots,n$, is any collection of
loops $\om^i_1,\ldots, \om^i_{m_i}$ on the dual complex whose links
are dual to links shared by pairs of triangles in $\tau$, i.e. the links
in the loops may be considered as connecting centres of neighbouring
triangles in $\tau$. By $|L_i|$ we denote the total number of links in
$\om^i_1,\ldots, \om^i_{m_i}$, i.e.\ their total length, and all
loops in $L_1,\ldots,L_n$ are required to be self-avoiding and pairwise
disjoint. Finally, $C_{\tau}(L_1,\ldots,L_n)$ denotes the order of the 
automorphism group
of $\tau$ with the configuration of loops $L_1,\ldots,L_n$. In figure~\ref{F0} we show a surface contributing to
$F_1(\underline{g})$.
 
Although in general 
$\tau$ does not consist of triangles alone we shall refer to $\tau$
as a triangulation.
In the following we shall be interested in the spherical limit
$N\rightarrow \infty$.

\newsection{Coarse graining \label{coarse}}
Given a polygonal loop $\gamma_0$ we let $T(\gamma_0)$
be the set of triangulations of planar topology bounded by $\gamma_0$
and define the one-loop function $W_{\gamma_0}(\underline{g})$ by
\beq
W_{\gamma_0}(\underline{g})=\sum_{\tau\in T(\gamma_0)}
\sum_{L_1,\ldots,L_n}
\prod_{j\geq 3}
(-g_j)^{N_j(\tau)}\left(\frac{g_0}{g_3}\right)^{|L_1|+\ldots
+|L_n|},
\label{oneloop}
\eeq
where $L_1,\ldots,L_n$ are as above.
Due to the relation
\beq
\sum_{j\geq 3}jN_j(\tau)=2L(\tau)-|\gamma_0|\;,
\label{sumrule}
\eeq
where $L(\tau)$ denotes the number of links in $\tau\in T(\gamma_0)$,
it follows that if $\gamma_0$ has length two, which we henceforth
assume,
then $W_{\gamma_0}(\underline{g})\equiv W(\underline{g})$ is obtained
by applying a suitable linear combination of the derivatives 
$g_j\frac{\partial }{\partial g_j}$, $j=0,3,\ldots$ to
$F_0(\underline{g})$.
\label{difop}
Applying a further differentiation yields the susceptibility
\beq
\chi_{00}(\underline{g}) =\sum_{\tau\in T(\gamma_0,\gamma')}
\sum_{L_1,\ldots,L_n}
\prod_{j\geq 3}
(-g_j)^{N_j(\tau)}\left(\frac{g_0}{g_3}\right)^{|L_1|+\ldots+|L_n|},
\label{chi00}
\eeq
where $T(\gamma_0,\gamma')$ denotes the set of triangulations of
cylindrical topology bounded by two loops $\gamma_0$ and $\gamma'$,
both of length two. 

It is well known that the sums~\rf{oneloop} and~\rf{chi00} are
convergent for $|g_j|$, $j=0,3,4,\ldots$, small enough and hence that
$W(\underline{g})$ and $\chi_{00}(\underline{g})$ are analytic in a
domain ${\cal A}$ of coupling constants. We shall be interested in the
singular behaviour of these quantities at the boundary $\partial {\cal
A}$ of the domain of analyticity, in the following referred to as the
critical surface. Assuming that the singular (or critical) behaviour
reflects properties of an underlying continuum surface theory we
expect the detailed structure at the scale of the lattice cutoff
of the discrete surfaces
(triangulations) that are summed over to be unimportant. In
particular, we may restrict the ensembles $T(\gamma_0)$ and
$T(\gamma_0,\gamma')$ to the corresponding ensembles
$\tilde{T}(\gamma_0)$ and $\tilde{T}(\gamma_0,\gamma')$ consisting of
surfaces which do not contain any loops of length two (except for the
boundary loops). In this way we define a one-loop function $\tilde{W}$
and a susceptibility $\tilde{\chi}_{00}$ in a domain of analyticity 
$\tilde{\cal A}$ and we expect by universality that there is a
one-to-one mapping between $\partial {\cal A}$ and $\partial
\tilde{{\cal A}}$ by which corresponding points represent identical
critical behaviour.

In order to exploit this, let us first express $W$ and
$\chi_{00}$ in terms of $\tilde{W}$ and
$\tilde{\chi}_{00}$. This is done simply by performing a partial
summation in~\rf{oneloop} and~\rf{chi00} over outgrowths on $\tau$
bounded by loops of length two as follows. Given a loop $\gamma$ of
length two in $\tau\in T(\gamma_0)$ there are two cases to consider:
\begin{enumerate}
\item
There is no loop in $L_1\cup\ldots \cup L_n$ which crosses
$\gamma$.
\item
A loop $\om^ i_j$ crosses $\gamma$, and since $\om^i_j$ is closed it 
actually crosses both links in $\gamma$. 
\end{enumerate}
By cutting $\tau$ along $\gamma$ the 
triangulation $\tau$ decomposes into a triangulation $\tau_0'\in 
T(\gamma_0,\gamma)$ and an outgrowth $\tau_1\in T(\gamma)$. In case 2 a 
segment of $\om^i_j$ is contained in $\tau_0'$ and the remaining part in 
$\tau_1$.  Both segments emerge from one link in $\gamma$ and end at the other
one.
Thus when summing over all outgrowths $\tau_1$ we obtain in case 2 a 
contribution $\left(1+W_1(\underline{g})\right)$ where $W_1(\underline{g})$
is given by the right hand side of equation~\rf{oneloop} with the sum 
restricted to
triangulations $\tau\in T(\gamma)$ whose boundary polygons are triangles 
with collections of self-avoiding and non-intersecting curves on the dual 
triangulation, all of which are closed except one which connects the two
boundary links of $\gamma$ and which has half-links at its ends, see figure~2.

\begin{figure}
\centerline{ \epsfbox{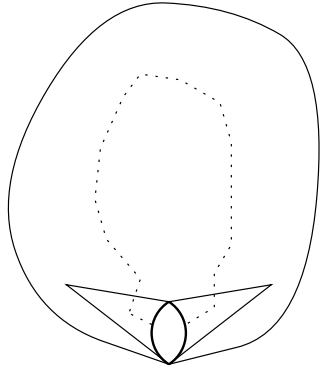}}
\caption{A surface contributing to $W_1(\underline{g})$}
\end{figure} 

By gluing the two links in the boundary component $\gamma$ in $\tau_0'$ 
together we obtain a triangulation $\tau_0\in T(\gamma_0)$ and the segment of 
$\om^i_j$ in $\tau_0'$ turns into a closed loop.
It follows that summation over maximal outgrowths allows us to write (see 
also~\cite{Dur94})
\bea
W(\underline{g})&=&\sum_{\tau\in\tilde{T}(\gamma_0)}
\sum_{L_1,\ldots,L_n}
\prod_{j\geq 3}(-g_j)^{N_j(\tau)}
\left(\frac{g_0}{g_3}\right)^{|L_1|+\ldots+|L_n|} \times 
\nonumber\\
&&
\left(1+W(\underline{g})\right)^{L(\tau)-\left(|L_1|+\ldots+|L_n|\right)-1}
\left(1+W_1(\underline{g})\right)^{|L_1|+\ldots+|L_n|},
\label{summation}
\eea
which as a consequence of~\rf{sumrule} can be rewritten 
as\footnote{For~\rf{wwtilde} to hold we actually need to 
distinguish the links in each
boundary two-loop. This convention, which we note
does not influence the singular behaviour of the
quantities under consideration, is the reason for the absence of symmetry
factors in~\rf{oneloop},~\rf{chi00} and~\rf{summation}.}
\beq
W(\underline{g})=\tilde{W}(\underline{\tilde{g}}),
\label{wwtilde}
\eeq
with
\bea
\tilde{g}_j&=&\left(1+W(\underline{g})\right)^{j/2}\, g_j
, \hspace{0.5cm} j\geq 3,
\label{ggtilde} \\
\tilde{g}_0&=&\left(1+W_1(\underline{g})\right)
\left(1+W(\underline{g})\right)^{1/2}
g_0.
\label{g0g0tilde}
\eea

The susceptibility may be treated similarly. Cutting along loops which 
decompose $\tau\in T(\gamma_0,\gamma)$ into two cylindrical triangulations 
yields a geometric series for the $2\times 2$ matrix 
$\chi_{ij}(\underline{g})$, $0\leq i,j\leq 1$, of susceptibilities defined as 
in~\rf{chi00} but with boundary conditions imposed at 
$\gamma_0$ and $\gamma$
as follows: For $\chi_{01}(\underline{g})=\chi_{10}(\underline{g})$ the 
summation in~\rf{chi00} is subject to the constraint that the boundary 
polygons in $\tau$ at, say, $\gamma_0$ are triangles and there is a polygonal 
curve in $L_1\cup\ldots\cup L_n$ connecting the two links in $\gamma_0$,
i.e. $L_1,\ldots,L_n$ are as in the definition of $W_1(\underline{g})$;
$\chi_{11}(\underline{g})$ is given by~\rf{chi00} with $\tau$ restricted  
so that its four boundary polygons are all triangles and in $L_1,\ldots,L_n$
there are two polygonal curves connecting two pairs of boundary links. We then
have
\beq
\chi(\underline{g})=\tilde{\chi}(\underline{\tilde{g}})\, \left (1-
\tilde{\chi}(\underline{\tilde{g}})\right)^{-1}
\label{chichitilde}
\eeq
where $1$ is the $2\times 2$ unit matrix.
Equations~\rf{wwtilde} and~\rf{chichitilde} are the desired expressions for 
$W(\underline{g})$ and $\chi(\underline{g})$ in terms of the corresponding 
quantities in the coarse grained tilde-model. At this point we can adopt the
arguments of~\cite{Dur94}. For the sake of completeness we shall briefly 
recapitulate them. 
First, let us note that the strongest divergence of the susceptibilities
$\chi_{ij}$ will allways be contained in $\chi_{00}$. The
susceptibilities $\chi_{01}$, $\chi_{10}$ and $\chi_{11}$ can be less
divergent than $\chi_{00}$, namely if the number of links in $L_1\cup
\ldots \cup L_n$ on the average occupies a vanishing fraction of the
total number of links in $\tau$, but the opposite situation can not occur.
We consider now a critical point 
$\underline{g}_c\in \partial {\cal A}$ at which
$\chi_{00}(\underline{g})$
 diverges 
with an associated critical exponent $\gamma>0$ such that
\beq
\chi_{00}(\underline{g})\sim \, |\underline{g}-\underline{g}_c|^{-\gamma}
\label{gamma+}
\eeq
as $\underline{g}$ approaches $\underline{g}_c$ transversally to 
$\partial {\cal A}$.
 As noted previously, 
$\chi_{00}(\underline{g})$ is obtained as a derivative of $W(\underline
g)$ (cf.\  page~\pageref{difop}).
 Since $W(\underline{g}_c)$ is finite as a consequence
of~\rf{wwtilde}, \rf{ggtilde} and~\rf{g0g0tilde} it follows that
\beq
W(\underline{g})\sim W(\underline{g}_c)+cst\cdot |\underline
g-\underline g_c|^{1-\gamma}.
\label{wgamma+}
\eeq
By diagonalising the susceptibility matrices in~\rf{chichitilde} it is seen 
that the divergence of $\chi(\underline{g})$ 
 is a consequence of the largest of the 
eigenvalues of $\tilde{\chi}(\underline{\tilde{g}})$ tending to 1 as 
$\underline{g}\rightarrow \underline{g_c}$. Thus assuming that this approach 
is given by a susceptibility exponent $\tilde{\gamma}\leq 0$ such that
\beq
\tilde{\chi}(\tilde{\underline{g}})\sim  \tilde{\chi}_c + C \cdot 
|\underline{\tilde{g}}-\underline{\tilde{g}}
(\underline{g}_c)|^{-\tilde{\gamma}} 
\label{chic}
\eeq
where $1-\tilde{\chi}_c$ is singular we obtain from~\rf{ggtilde}--\rf{wgamma+}
that
\beq
|\tilde{\underline{g}}-\underline{\tilde{g}}(\underline{g}_c)|^{-\gamma}
\sim |\underline{\tilde{g}}-\underline{\tilde{g}}(\underline{g}_c)|^
{\tilde{\gamma}\,\left(1-\gamma\right)} 
\eeq
i.e.
\beq
\gamma=\frac{\tilde{\gamma}}{\tilde{\gamma}-1}.
\label{gamma+-}
\eeq
Alternatively, it may happen that $\underline{\tilde{g}}(\underline{g}_c) \in
\tilde{{\cal A}}$ i.e.\ that $\tilde{\chi}$ is analytic at 
$\tilde{\underline{g}}(\underline{g}_c)$. This case corresponds to
$\tilde{\gamma}=-1$ in~\rf{chic}, i.e.
\beq
\gamma=\frac{1}{2}.
\label{gammabp}
\eeq
Returning to the universality assumption discussed previously we have thus 
obtained that if there exists a critical point $\underline{g}_c$ with 
susceptibility exponent $\gamma >0$ then either $\gamma=\frac{1}{2}$ or there
exists a corresponding critical point with critical exponent $\tilde{\gamma}$
related to $\gamma$ by~\rf{gamma+-}. In section~\ref{critical}  we show by 
exploiting the exact solution of the model that both cases~\rf{gamma+-} and
\rf{gammabp} are actually realized for $n>2$, and that by suitable choices of 
the potential, $\tilde{\gamma}$ may assume the values $-\frac{1}{m}$,
$m=2,3,\ldots,$ with corresponding values of $\gamma=\frac{1}{m+1}$.

\newsection{The exact solution \label{solution} }
For the following considerations it is convenient to write the partition
function of the model in the form
\beq
Z=e^{N^2 F}=\int_{N\times N} dM \prod_{i=1}^n dA_i \exp \left\{-N\tr 
\left( V(M)+\sum_{i=1}^nMA_i^2\right)\right\}
\label{partition}
\eeq
where $V(M)$ is an arbitrary polynomial potential,
$V(M)=\sum_{j=1}g_j/j\, M^j$. From~\rf{partition} the form of~\rf{Z} is
easily recovered by a linear shift of $M$.
 Expressed in terms
of the eigenvalues, $\{\lambda_i\}$, of the matrix $M$ the 
integral~\rf{partition}
reads
\beq
Z\propto \int \prod_{i=1}^N
d\lambda_i\exp\left\{-N\sum_jV\left(\lambda_j\right)\right\} 
\prod_{j<k}(\lambda_j-\lambda_k)^2\prod_{i,l}
(\lambda_i+\lambda_l)^{-n/2}.
\label{eigenvalues} 
\eeq
In the large $N$ limit the eigenvalue configuration is determined by the
saddle point of the integral above and the eigenvalues are confined to a
compact region of the real axis~\cite{BIPZ78}. 
In the following we consider the situation where the eigenvalues live
on only one  interval $[a,b]$ on the positive real axis. 
We shall make use of the
exact solution of the model found in this situation in~\cite{EK95,EK96}.

Let us briefly
recall the main characteristics of this solution.
It proved convenient to parametrize the solution 
 in terms of two sets of
basis functions,
$\left\{G_a^{(k)}(p),G_b^{(k)}(p)\right\}$  and
$\left\{\tilde{G}_a^{(k)}(p),\tilde{G}_b^{(k)}(p)\right\}$, $k\in
\{0,1,\ldots \}$. 
The $G$-functions 
 are
defined by certain requirements on their analyticity structure and
their asymptotic behaviour. The $\tilde{G}$-functions 
 are obtained from the $G$-functions by
the replacement $\nu\rightarrow (1-\nu)$. Furthermore, functions with
subscript $b$ are obtained from functions with subscript $a$ by the
interchangement $a\leftrightarrow b$.
 Explicit expressions for all basis functions in
terms of $\theta$-functions can be written down. In the following we
shall not need the explicit expressions for the $G$- and
$\tilde{G}$-functions, only the fact that they fulfill certain
recursion relations. For instance it holds that
\bea
\frac{\partial G_a^{(k)}(p)}{\partial
a^2}&=&\lambda_a^{(k)}G_a^{(k+1)}(p),
\label{diff}\\
pG_a^{(k)}(p)&=&\tilde{G}_a^{(k-1)}(p)+s_a^{(k)}\tilde{G}_a^{(k)}(p)
\label{tilde}
\eea
where $\{\lambda_a^{(k)}\}$ and $\{s_a^{(k)}\}$ are some (finite)
constants. These
constants can be expressed entirely in terms of the endpoints of the
support of the eigenvalue distribution, $a$
and $b$, and two additional parameters $e$ and $\alpha$. The
parameters
$e$ and $\alpha$ are given by
\bea
e&=&a\,\mbox{sn} \left(i(1-\nu)K',k\right) \\
\alpha&=&b\left\{Z\left(i(1-\nu)K',k\right)+i(1-\nu)\frac{\pi}{2K}\right\},
\hspace{0.7cm} k=\frac{a}{b}
\label{alpha}
\eea
where $Z$ is the Jacobian zeta function and $K$ and $K'$ are the
complete elliptic integrals of the first kind.
The quantity $\alpha$ is essentially a first derivative of $e$,
\beq
(b^2-a^2)\,\rho_a=\frac{\overline{\sqrt{e}}}{e}\alpha+b^2-e^2
\label{rho}
\eeq
where
\beq
\rho_a=\frac{a^2}{e^2}\frac{\partial e^2}{\partial a^2},\hspace{1.0cm}
\overline{\sqrt{e}}=\sqrt{(e^2-a^2)(e^2-b^2)}.
\eeq
Obviously, recursion relations similar to~\rf{diff} and~\rf{tilde} hold
in the case where $a$ is replaced by $b$. Furthermore similar
recursion relations hold when untilded functions and constants
are replaced by
their tilded analogues. 
In this connection it is useful to note that
\beq
\tilde{\rho}_a=1-\rho_a=\rho_b,
\hspace{1.0cm}\tilde{e}=-\frac{ab}{e},
\hspace{1.0cm}
\frac{\overline{\sqrt{\tilde{e}}}}{\tilde{e}}=
\frac{\overline{\sqrt{e}}}{e}.
\label{ttilde}
\eeq
In reference~\cite{EK95} the constants $\lambda_a^{(0)}$,
$\lambda_a^{(1)}$ and $s_a^{(1)}$ were determined and recursion
relations which allow one to extract from these $\lambda_a^{(k)}$ and 
$s_a^{(k)}$ for all $k$, were derived. From this analysis we shall need
the following results:
\beq
\lambda_a^{(0)}=-\frac{1}{2}\,i\,\tan \left(\frac{\nu \pi}{2}\right)
\frac{e^2}{a^2}\frac{e^2-a^2}{e\overline{\sqrt{e}}}\rho_a,
\label{lambda}
\eeq
\beq
s_a^{(1)}=\frac{1}{2}\left(\frac{1-\rho_a}{\lambda_a^{(0)}}\right),
\hspace{1.0cm}
s_a^{(k+1)}=s_a^{(k)}\,\frac{\tilde{\lambda}_a^{(k)}}{\lambda_a^{(k)}}.
\label{srelation}
\eeq
The quantities mentioned so far depend on the matrix model potential
only implicitly, namely via $a$ and $b$. All
explicit dependence on the potential $V(M)$
is described via a set of moment variables defined by
\beq
M_k=\oint_C\frac{d\om}{2\pi i}V'(\om)\tilde{G}_a^{(k)}(\om),
\hspace{1.0cm} J_k=M_k(a\leftrightarrow b)
\label{momdef}
\eeq
where $C$ is a curve which encircles the interval $[a,b]$.
In particular,
the equations which determine $a$ and $b$
can conveniently be expressed in terms of moment variables. They read
\bea
M_0&=&\oint_C \frac{d\om}{2\pi i}V'(\om)\tilde{G}^{(0)}(\om)=0,
\label{B1}\\
M_{-1}&=&\oint_C\frac{d\om}{2\pi i}V'(\om)\om G^{(0)}(\om)=2-n
\label{B2}
\eea
where $G^{(0)}(p)\equiv G_a^{(0)}(p)=G_b^{(0)}(p)$.

\newsection{Critical behaviour for $n>2$ \label{critical} }

We now turn to discussing the critical properties of the
model. For that purpose we introduce an overall coupling constant in
front of our potential, i.e.\ we replace $V(M)$ by $V(M)/T$ where $T$
is to be thought of as the cosmological constant. Then we define the string
susceptibility $U(T)$ by
\beq
U(T)=\frac{d^2}{dT^2}\left(T^2F_0\right)
\eeq
where $F_0$ is the genus zero contribution to the free energy. 
The leading singularity of $U(T)$ is the same as that of the
susceptibility $\chi_{00}(\underline{g})$, introduced in the previous section.
In geometrical terms $U(T)$ expresses the average number of links minus
the average area for surfaces with one boundary component, i.e.
\beq
U(T)\sim \langle L(\tau)-A(\tau) \rangle_{\tau\in T_0(\gamma)}
\eeq
where $A(\tau)=\sum_{j\geq 3}N_j(\tau)$.
We note that in the vicinity of a singular point the quantities
$\langle L(\tau)\rangle$, $\langle A(\tau)\rangle$ and $U(T)$ 
obey the same scaling behaviour.\label{area}

What we
will aim at calculating is, in case the model has a critical point at
$T=T_c$, the value of the critical index $\gamma$ 
\beq
U(T)\sim\left(T_c-T\right)^{-\gamma}.
\eeq
In reference~\cite{EK96} the following expression for $dU(T)/dT$ was
derived
\beq
\frac{dU(T)}{dT}=\left(1-\frac{n}{2}\right)\frac{1}{b^2-a^2}\left\{
\frac{b^2-\tilde{e}^2}{\tilde{e}^2}\frac{da^2}{dT}
-\frac{a^2-\tilde{e}^2}{\tilde{e}^2}\frac{db^2}{dT}\right\}.
\label{dUdT}
\eeq
This expression is universal in the sense that it does not
contain any direct reference to the matrix model potential. We can
make the expression~\rf{dUdT} even more explicit by deriving closed expressions
for $da^2/dT$ and $db^2/dT$. To do so we first differentiate the
boundary equation~\rf{B1} with respect to $T$. Using the
relation~\rf{diff} and the definition of the moment
variables~\rf{momdef} we get
\beq
\tilde{\lambda}_a^{(0)}\,M_1\, \frac{da^2}{dT}+
\tilde{\lambda}_b^{(0)}\,J_1\, \frac{db^2}{dT}=0.
\label{diff1}
\eeq
Next we differentiate the boundary equation~\rf{B2} with respect to
$T$. This gives
\beq
\frac{1}{2}(1-\rho_a)\,M_1 \, \frac{da^2}{dT}+
\frac{1}{2}(1-\rho_b)\,J_1 \, \frac{db^2}{dT}=2-n
\label{diff2}
\eeq
where on the way we have made use of the relations~\rf{diff},
\rf{tilde}  and~\rf{srelation}. Combining~\rf{diff1} and~\rf{diff2} and
using~\rf{lambda} we
get
\beq
\frac{da^2}{dT}=2(2-n)\,
\frac{\left(b^2-\tilde{e}^2\right) a^2}{(b^2-a^2)\tilde{e}^2}\,
\frac{1}{(1-\rho_a)M_1},
\label{dadt}
\eeq
and similarly  $db^2/dT$ equals the right hand side of~\rf{dadt} with $a$ and
$b$ interchanged.
Inserting the expressions for $da^2/dT$ and $db^2/dT$ into~\rf{dUdT}
we get
\beq
\frac{dU(T)}{dT}=(2-n)^2\frac{1}{(b^2-a^2)^2}\left\{
\frac{a^2(b^2-\tilde{e}^2)^2}{\tilde{e}^4}\frac{1}{(1-\rho_a)M_1}
-\frac{b^2(a^2-\tilde{e}^2)^2}{\tilde{e}^4}\frac{1}{\rho_a J_1}\right\}.
\label{dUdTfin}
\eeq
We stress that the expression~\rf{dUdTfin} is valid for any value of
$n$ and any potential $V(M)$. In formula~\rf{dUdTfin} it is easy to spot
singular points of the model. From now on we shall concentrate
 our analysis on the
case $n>2$. Accordingly we set $\nu=i\bar{\nu}$ with $\bar{\nu}$
real. We hence have
\beq
n=2\cosh (\bar{\nu}\pi).
\eeq
Obviously a singularity appears if $\tilde{e}$ becomes equal to
zero. This happens whenever $k=\frac{a}{b}$ takes one of the values for
which
\beq
\bar{\nu}K'=2mK,\hspace{0.7cm} m=1,2,\ldots
\eeq
As argued in reference~\cite{EK96} the model only has meaning until
the first of these singularities is reached i.e. for $k>k_c$ where
$k_c$ is given by
\beq
\frac{K'}{K}(k_c)=\frac{2}{\bar{\nu}}.
\eeq
We note that the critical value of $k$ corresponds to a finite 
(positive) value
of $a$. In particular, this means that for $n>2$ the model makes no
sense at the point $a=0$ which would be the naive analytical
continuation of the critical point for $n\in [-2,2]$ to $n>2$. In
addition to the singularity which occurs for $\tilde{e}=0$ we might
expect to
encounter a singularity if $1-\rho_a=0$ or $\rho_a=0$. However, as
shown in reference~\cite{EK96} the first possibility can be
realized only for $k<k_c$ and the second one does not imply any kind
of non analytical behaviour. 
Let us also note, that the parameter $\tilde{e}$ can not diverge when $n>2$.
It is possible for $\rho_a$ to diverge but only at the points $\tilde{e}=0$.
Finally the model can become singular if one of
the moments $M_1$ (or $J_1$) acquires a zero of some order. 
It follows immediately from the analysis of~\cite{EK96} that at the
simplest possible of these points, $k=k_c$, $M_1^c\neq 0$, $J_1^c\neq 0$,
the model is in a branched polymer phase, i.e.
\beq
\gamma=\frac{1}{2}, \hspace{0.7cm} k=k_c,\mbox{ }
 M_1^c\neq 0,\mbox { } J_1^c \neq 0.
\eeq
We shall now
argue that the remaining critical points for $n>2$ can be classified in
the following way
\begin{itemize}
\item
$M_1^c=M_2^c=\ldots=M_{m-1}^c=0,\hspace{0.3cm} M_m^c\neq 0,
\hspace{0.3cm} k>k_c$\\
$\gamma=-\frac{1}{m}$.
\item
$M_1^c=M_2^c=\ldots=M_{m-1}^c=0,\hspace{0.3cm} M_m^c\neq 0,
\hspace{0.3cm} k=k_c$ \\ 
$\gamma = +\frac{1}{m+1}$.
\end{itemize}
In other words we have exactly the situation predicted by the coarse
graining argument: By fine tuning the potential we can reach 
critical points with positive values of the string susceptibility
exponent and these points have dual partners where the string
susceptibility exponent is negative. It appears that the possible
values for the dual pairs of exponents are
$\left(\tilde{\gamma},\gamma\right)=\left(-\frac{1}{m},\frac{1}{m+1}\right)$,
$m=2,3,\ldots$ and 
that the critical points at which $\gamma=\frac{1}{m+1}$, $m=2,3,\ldots$
are located where the critical surface corresponding to a $c<1$ theory
($\tilde{\gamma}=-\frac{1}{m}$) intersects the critical surface
 corresponding to
branched polymer behaviour.

For simplicity, let us first
consider a critical point corresponding to $M_1^c=0$,
$M_2^c\neq 0$, $k>k_c$ and let us approach the point by fixing
the coupling constants of $V(M)$ at their critical values and letting
$T$ approach $T_c=1$. The boundary equations~\rf{B1} and~\rf{B2}  
tell us how $a$ and $b$
approach their critical values $[a_c,b_c]$ under this fine tuning.
Expanding equation~\rf{B1} in powers of
$(b^2-b_c^2)$ and $(a^2-a_c^2)$ and keeping only leading order terms
we get
\beq
\tilde{\lambda}_b^{(0)}J_1^c\,(b^2-b_c^2)+\frac{1}{2}
\tilde{\lambda}_a^{(0)}\tilde{\lambda}_a^{(1)}M_2^c\,(a^2-a_c^2)^2=0
\eeq
where we have made use of the relation~\rf{diff} and exploited the
fact that $M_1^c=0$ while $M_2^c\neq 0$. This gives
\beq
(b^2-b_c^2)=-\frac{1}{2}\frac{\tilde{\lambda}_a^{(0)}\tilde{\lambda}_a^{(1)}}
{\tilde{\lambda}_b^{(0)}} \frac{M_2^c}{J_1^c}\,(a^2-a_c^2)^2
\label{ba}
\eeq
It is easy to see that $J_1$ must necessarily be different from $M_1$ and
hence non-vanishing. 
Treating the second boundary
equation~\rf{B2} in a similar manner and taking into account the
relations~\rf{diff} and~\rf{tilde} one finds
\beq
\lambda_b^{(0)}s_b^{(1)}J_1^c\,\left(b^2-b_c^2\right)+
\frac{1}{2}\lambda_a^{(0)}\lambda_a^{(1)}s_a^{(2)}M_2^c
 \left(a^2-a_c^2\right)^2
=(2-n)\left(T-T_c\right)
\label{B2c}
\eeq
Inserting the expression~\rf{ba} for $(b^2-b_c^2)$ in~\rf{B2c} and
making use of~\rf{lambda} and~\rf{srelation} we finally get
\beq
\left.\frac{1}{4}M_2^c\,\tilde{\lambda}_a^{(1)}\tilde{\rho_a}\,
\left\{
\frac{a^2-b^2}{\tilde{e}^2-b^2}\right\}\,\frac{\tilde{e}^2}{a^2}
\right|_{\,T=T_c}
\left(a^2-a_c^2\right)^2=(n-2)\left(T_c-T\right)
\label{aT}
\eeq
Since per assumption $\tilde{e}\neq 0$ ($k>k_c$)
and  $|\tilde{e}|<\infty$ as well as $\tilde{e}^2\neq b^2$ ($n>2$) 
we have
\beq
(a^2-a_c^2)^2\sim (T_c-T)
\label{a2T}
\eeq
Furthermore, expanding $M_1$ around $a=a_c$ and exploiting the fact
that $M_2\neq 0$ we get
\beq
M_1\sim (a^2-a_c^2)
\label{M1scal}
\eeq
Finally it follows from~\rf{alpha} and~\rf{rho} that $\rho_a$ stays
finite as $a\rightarrow a_c$. In conclusion we find
\beq
\frac{dU(T)}{dT}\sim (a^2-a_c^2)\sim (T_c-T)^{-1/2}
\eeq
which implies that
\beq
\gamma=-\frac{1}{2}.
\eeq
It is obvious how the argument goes in the general case. For
$M_1^c=M_2^c=\ldots=M_{m-1}^c=0$, $M_m^c \neq 0$ and $k>k_c$ one gets
\beq
M_1\sim (a^2-a_c^2)^{m-1}, \hspace{1.0cm}
 (a^2-a_c^2)^m \sim (T_c-T)
\eeq
which gives
\beq
\frac{dU(T)}{dT}\sim (T-T_c)^{-\frac{m-1}{m}},
\hspace{1.0cm}\mbox{i.e.}\hspace{0.7cm} \gamma=-\frac{1}{m}
\eeq

Now let us turn to considering the type of critical point where in
addition $\tilde{e}=0$. For simplicity, let us assume that otherwise
we have the same situation as before, i.e. $M_1^c=0$, $M_2^c\neq
0$. In that case, obviously the relation~\rf{ba} between $(b^2-b_c^2)$
and $(a^2-a_c^2)$ remains the same. So does the relation~\rf{aT} but
we see that now the coefficient of $(a^2-a_c^2)^2$ becomes equal to
zero:
From~\rf{alpha} and~\rf{rho} (expressed in terms of tilded variables instead of
untilded ones) we conclude that $\tilde{\rho}_a\cdot \tilde{e}$ is
finite as $k\rightarrow k_c$. This follows from the fact that
$\tilde{\alpha}$ is finite for $k=k_c$. Hence the total coefficient
vanishes.
This implies that the relation~\rf{a2T} is replaced by
\beq
(a^2-a_c^2)^3\sim (T_c-T).
\eeq
(It is easy to convince oneself that under the given circumstances the
coefficient of the term cubic in $(a^2-a_c^2)$ in the expansion
of~\rf{B2} can not vanish.) 
For the scaling of $M_1$ we still have the same behaviour as
in~\rf{M1scal}. However, in the present case, in addition $\tilde{e}$
and $\tilde{\rho}_a$ will scale.  As mentioned above $\rho_a \cdot
\tilde{e}$ is finite as $k\rightarrow k_c$. Now, expanding
$\tilde{e}$ around its critical value we get
\beq
\tilde{e}=
\left.\frac{1}{2}\frac{\tilde{e}}{a^2}\tilde{\rho}_a\right|_{k=k_c}
\left(a^2-a_c^2\right)
+\left.\frac{1}{2}\frac{\tilde{e}}{b^2}(1-\tilde{\rho}_a)\right|_{k=k_c}
\left(b^2-b_c^2\right)
\eeq
from which we conclude that
\beq
\tilde{e}\sim (a^2-a_c^2)\hspace{0.7cm}
\mbox{as}\hspace{0.7cm} k\rightarrow k_c.
\eeq
In total we get from~\rf{dUdTfin}
\beq
\frac{dU(T)}{dT}\sim \frac{1}{\left(a^2-a_c^2\right)^4}\sim
\left(T_c-T\right)^{-4/3}
\eeq
or
\beq
\gamma=+\frac{1}{3}.
\eeq
Also for the critical points at which $\tilde{e}=0$ it is obvious how
the scaling will be in the general case
$M_1^c=M_2^c=\ldots=M_{m-1}^c=0$, $M_m^c\neq 0$. One finds
\beq
M_1\sim (a^2-a_c^2)^{m-1}, \hspace{1.0cm} (a^2-a_c^2)^{m+1} 
\sim (T_c-T)
\eeq
whereas the scaling of $\tilde{e}$ and $\tilde{\rho_a}$ is
unchanged. For $\frac{dU(T)}{dT}$ one hence has
\beq
\frac{dU(T)}{dT}
\sim (a^2-a_c^2)^{-m-2}
\sim \left(T_c-T\right)^{-1-\frac{1}{m+1}}
\eeq
which implies
\beq
\gamma=+\frac{1}{m+1}.
\eeq 

\newsection{The average number of loops \label{loop} }

In this section we calculate a geometric observable which shows
strikingly different behaviour at the critical points with
respectively positive and negative values of $\gamma$. To begin with,
let us note that the dependence on $n$ in sums like~\rf{Fh},
\rf{oneloop} and~\rf{chi00} can easily be explicitly exposed by first
summing over configurations of (uncoloured) loops and afterwards
summing over colours. For instance for $F_0(\underline{g})$ we have
\beq
F_0(\underline{g})=\sum_{\tau\in T_0}
\sum_{\{ {\cal L}\}}C_{\tau}^{-1}({\cal L})\prod_{j\geq 3}(-g_j)^{N_j(\tau)}
\left(\frac{g_0}{g_3}\right)^{|{\cal L}|}
n^{{\cal N}_{\cal L}}
\label{F0a}
\eeq
where the notation is as in~\rf{Fh} except that $\{ {\cal L} \}$ now
refers to a given configuration of loops. By ${\cal N}_{\cal L}$ we
mean the total number of loops in $\{ {\cal L}\}$ and by $| {\cal L}|$
the total number of links of the loops in $\{{\cal L}\}$. Since
$F_0(\underline{g})$ is finite at all critical points of the model,
calculating $\frac{d F_0(\underline{g})}{d n}$ will give
us the average value of ${\cal N}_{\cal L}$ up to a finite
normalization constant. In particular, if the quantity $\frac{d
F_0(\underline{g})}{d n}$ diverges it means that the average
number of loops on the surfaces contributing to the sum~\rf{F0a}
diverges. The same argument applies if $F_0(\underline{g})$ is replaced
by $W_{\gamma_0}(\underline{g})$ or any other correlation function
which itself is finite at the critical point. In the following we
will use the notation of equation~\rf{partition} and the quantity we aim
at calculating is $\frac{d}{d n}\langle \frac{1}{N}\tr
M\rangle$. First we note that due to~\rf{eigenvalues} we have
\bea
\frac{d F}{d n}&=&
-\frac{1}{2}\frac{1}{Z}\frac{1}{N^2}\int 
\prod_j d\lambda_j \exp \left\{-N\sum_iV(\lambda_i)\right\}
\prod_{i<j}\left(\lambda_i-\lambda_j\right)^2 \times \nonumber \\
&& 
\prod_{i,j}\left(\lambda_i+\lambda_j\right)^{-n/2}\sum_{i,j}\log
(\lambda_i+\lambda_j) \nonumber\\
&=& -\frac{1}{2}\left\langle
\frac{1}{N^2}\sum_{i,j}\log(\lambda_i+\lambda_j)\right\rangle. 
\eea
In the large $N$ limit we can write this as
\beq
\frac{d F}{d n} =-\frac{1}{2}
\oint \frac{d\om_1}{2\pi i}\oint \frac{d \om_2}{2\pi i}\,
W(\om_1)W(\om_2)\log(\om_1+\om_2)
\label{dFdn}
\eeq
where we have gone through the steps of reference~\cite{BIPZ78} of
introducing a density of eigenvalues and a corresponding generating
functional of expectation values
$W(p)=\left\langle\frac{1}{N}\tr\frac{1}{p-M}\right\rangle$. The contours
in~\rf{dFdn} encircle the cut of $W(p)$ (and not that of $W(-p)$)
or equivalently the
support of the eigenvalue distribution (and not its mirror image with respect
to zero). Now let us act on both sides
of~\rf{dFdn} with the loop insertion operator $d/dV(p)$ defined
by~\cite{AJM90}
\beq
\frac{d}{dV(p)}=-\sum_{j=1}^{\infty} \frac{j}{p^{j+1}}
\frac{d}{d g_j}.
\eeq
This gives
\beq
\frac{d W(p)}{d n}=
-\oint\frac{d\om_1}{2\pi i}\oint\frac{d\om_2}{2\pi
i}W(\om_2,p)W(\om_1)\log(\om_1+\om_2) 
\eeq
where
$W(\om,p)=
\left\langle
\tr\frac{1}{\om-M}\tr\frac{1}{p-M}\right\rangle_{conn}$.
In~\cite{EK95} a closed expression for $W(\om,p)$ has been obtained in the
form 
\beq
W(\om,p)=\frac{\partial}{\partial \om}\left\{H(\om,p)\right\}
\eeq
with
\bea
H(\om,p)&=&\frac{1}{4-n^2}\left\{\frac{1}{p^2-\om^2}\left(
ip\tilde{G}(p)\left[i(w^2-e^2)\tilde{G}(\om)-
\om \frac{\overline{\sqrt{e}}}{e}G(\om)\right] \nonumber \right.\right.\\
&& \left.\left.-G(p)\left[\left(w^2-\tilde{e}^2\right)\om G(\om)+i\om^2
\frac{\overline{\sqrt{e}}}{e}\tilde{G}(\om)\right]\right)
-n\frac{1}{p+\om}-2\frac{1}{p-\om}\right\} \nonumber
\label{H}
\eea
Hence we can write
\beq
\frac{dW(p)}{dn}=
-\oint \frac{d\om}{2\pi i}W(-\om)H(\om,p).
\eeq
Taking the coefficients of the $\frac{1}{p^2}$ terms in this equation
gives
\beq
\frac{d}{d n}\langle \frac{1}{N}\tr M\rangle=
-\frac{2i\sin\left(\frac{\nu\pi}{2}\right)}{4-n^2}
\oint\frac{d\om}{2\pi i}W(-\om)\left[i(w^2-e^2)\tilde{G}(\om)-
\om\frac{\overline{\sqrt{e}}}{e}G(\om)\right]
\label{dn}
\eeq
Let us now consider the quantity 
$\frac{d}{dn}\left\langle \frac{1}{N}\tr M\right\rangle$
in the vicinity of a critical point with $\gamma>0$. 
In order to determine the possible divergence of the integrand it is
necessary to write down the various contributions in terms of
$\theta$-functions exploiting the explicit expression for $G(\om)$
found in reference~\cite{EK96}.  The factor $e^2$ diverges
as $\frac{1}{\tilde{e}^2}$ and the factor $\frac{\overline{\sqrt{e}}}{e}$ as
$\frac{1}{\tilde{e}}$ (cf.\ Eq.~\rf{ttilde}). The function
$\tilde{G}(\om)$ is finite in the scaling limit whereas the function
$G(\om)$ contains a term which diverges as $\frac{1}{\tilde{e}}$. A careful
analysis shows that the leading singularities of the two terms in square
brackets in~\rf{dn} cancel, so that their contribution to the integrand
  diverges not
as $\frac{1}{\tilde{e}^2}$ but as $\frac{1}{\tilde{e}}$, i.e.\ as
$(a^2-a_c^2)^{-1}$ (cf.\ section~\ref{critical}). The one-loop
correlator, $W(p)$, although parametrized in terms of the divergent
function $G(p)$, is finite as it should be. This can again be seen by
exploiting the explicit expression for $G(p)$ in terms of
$\theta$-functions. In addition, at the critical points under
consideration, the cut of $W(-p)$ remains at a finite distance from
the cut of $W(p)$ and hence $W(-p)$ does not give rise to 
non-analytical behaviour of the integral. In conclusion, we have
\beq
\left\langle {\cal N}_{\cal L}\right\rangle\sim \frac{1}{\tilde{e}}
 \sim (a^2-a_c^2)^{-1},
\hspace{0.7cm} \gamma>0.
\label{NL}
\eeq
The existence of critical points for $n>2$ with $\left\langle {\cal N}_{\cal
L}\right\rangle $ diverging was conjectured in~\cite{Kosrev}.
The average area of surfaces with one boundary component obeys
the following scaling relation (cf.\ page~\pageref{area})
\beq
\left \langle A \right \rangle\sim (T_c-T)^{-\gamma}\sim (a^2-a_c^2)^{-1},
\hspace{0.7cm} \gamma >0.
\label{A1}
\eeq
We see that the divergence of the area of surfaces with one
boundary component is of the same
order as the divergence of the average number of loops. In order 
 for~\rf{NL} and~\rf{A1} to be consistent the average length of a single
loop must be finite.

\newsection{Discussion \label{discussion}}

For the case $n\in [-2,2]$
the phase space structure of the $O(n)$ model on a random lattice 
has
been understood since the late eighties~\cite{Kos89,DK88,KS92,EZ92}. 
For this range of $n$ values there exists a series of
critical surfaces where the model exhibits the scaling behaviour of
pure 2D gravity and its multi-critical versions, i.e.\  2D gravity
coupled to minimal conformal field theories of the type $(2,2m-1)$. At
these points $\gamma$ takes one of the values $\gamma=-\frac{1}{m}$,
$m=2,3,\ldots$. In addition, there exists a critical surface at which
$\gamma=-\frac{\nu}{1-\nu}$, $\nu$ being related to $n$ by
$n=2\cos(\nu \pi)$.
 When the critical surface corresponding to
$\gamma=-\frac{1}{m}$ intersects the critical surface corresponding to
$\gamma=-\frac{\nu}{1-\nu}$, new critical behaviour emerges and
$\gamma$ changes to
\beq
\gamma=\frac{-2\nu}{m-\eta_{m+1}+1-\nu},\hspace{0.8cm}
\eta_{2k}=\nu, \hspace{0.4cm} \eta_{2k+1}=1-\nu.
\eeq
 For $\nu$ rational the corresponding continuum theory describes
(a sector of) a minimal conformal model coupled to 2D quantum gravity
and by a suitable choice of $\nu$ and of the potential of the model
one may reach any minimal conformal model.
We note that the string susceptibility is always finite when $n\in ]-2,2]$.

 For $n>2$ the critical surfaces corresponding to $\gamma=-\frac{1}{m}$,
 $m=2,3,\ldots$ are still present. However, in this case the
singularity
corresponding to $\gamma=-\frac{\nu}{1-\nu}$ is prevented by the
occurrence of
 a new type of
 singularity at which the value of $\gamma$ is $+\frac{1}{2}$.
It is interesting to note that
 this value is independent of $n$. In analogy with what was the case
 for $n\in[-2,2]$ when a critical surface corresponding to
 $\gamma=-\frac{1}{m}$, $m=2,3,\ldots$ intersects the critical surface
 corresponding to $\gamma=\frac{1}{2}$ a new type of singular
 behaviour is seen. Here the associated value of the string
 susceptibility exponent is $\gamma=\frac{1}{m+1}$. As opposed to the
 case $n\in [-2,2]$ we have for $n>2$ critical points at which the
 string susceptibility diverges. This divergence can be traced back to a
 divergence of the average number of loops, $\langle {\cal N}_{\cal
 L}\rangle$, on surfaces with one boundary component.

Let us mention that special cases of the phase structure found here for
the $O(n)$ model for $n>2$ has been observed in
models constructed by adding to the usual one-matrix model
interaction terms of the type $(\tr M^2)^2$ or $(\tr
 M^4)^2$~\cite{curvature}. These
models, however, do not as the $O(n)$ model contain all the minimal
conformal models coupled to gravity  and do not have a regular lattice
formulation. Hence their interpretation as models describing 2D
gravity interacting with matter is less clear. It is not surprising,
however, that these models belong to the same universality class as
the $O(n)$ model on a random lattice for some range of $n$ values,
since the integral~\rf{Z} can be rewritten as
\beq
Z=\int dM \exp\left\{-N\tr V(M)+\frac{n}{2}\sum_{m+p\geq 1}^{\infty}
\frac{g_0^{m+p}}{m+p} \frac{(m+p)!}{m!\, p!}\tr M^m \tr M^p\right\}
\eeq

We finally note that in analogy with the multicritical models with $c<1$, the
possible continuum limits associated with critical points at which
$\gamma=\frac{1}{m}$, $m=4,5\ldots$, are expected to be non-unitary, whereas
for $\gamma=\frac{1}{3}$  unitarity should hold. Clearly, the detailed
properties of these theories constitute an interesting issue.

\end{document}